\documentclass[12pt]{article}
\usepackage[xdvi]{graphicx}
\usepackage{amssymb}
\usepackage{color}
\newcommand{\bea}{\begin{eqnarray}}
\newcommand{\eea}{\end{eqnarray}}

\makeatletter
\@addtoreset{equation}{section}
\makeatother

\def\ignore#1{{}}

\begin{document}
\begin{titlepage}
\begin{flushright}
OU-HET 671/2010
\end{flushright}

\vspace{25ex}

\begin{center}
{\Large\bf 
Intermediate-scale vertex corrections \\
\vspace{1ex}
for zero mode in warped space
}
\end{center}

\vspace{1ex}

\begin{center}
{\large
Nobuhiro Uekusa
}
\end{center}
\begin{center}
{\it Department of Physics, 
Osaka University \\
Toyonaka, Osaka 560-0043
Japan} \\
\textit{E-mail}: uekusa@het.phys.sci.osaka-u.ac.jp
\end{center}


\vspace{3ex}

\begin{abstract}

Non-Abelian gauge theory 
with a warped extra dimension
is studied
as a quantum field theory at an intermediate scale
that is regarded as being much lower than the scale of
the geometry stabilization and the Planck scale.
Loop corrections for zero-mode vertices
are diagrammatically
calculable in perturbation at the intermediate
scale.
The contribution for each diagram
can be compared to the correspondent
in the four-dimensional theory.
It is found that for a part of the contributions
the coefficient for the logarithmic scaling
has the same value as the four-dimensional results.
A viewpoint to 
treat corrections associated with
higher-dimensional operators 
is also discussed.

\end{abstract}

\end{titlepage}


\newpage

\section{Introduction}

Physics of extra dimensions 
is an interesting possibility 
of physics
beyond the standard model~\cite{Manton:1979kb}-%
\cite{Appelquist:2000nn}.
As in a picture believed commonly, 
physical quantities are momentum-dependent
so that quantum corrections should be considered.
However, it is far from the
full understanding
when extra dimensions are
included.
There are problems of higher-dimensional theory.
One arises from a property of non-renormalizability.
This inevitably gives rise to 
innumerable higher-dimensional operators.
Another is that loop integrals of virtual processes 
have higher degrees of divergence than the corresponding
four-dimensional integrals.
These problems are related to 
applicability 
of the theory to describe high-energy behavior.
It may be 
prospective to claim that
quantum effects in higher-dimensional theory 
are derived only in an ultraviolet completion.
As such a circumstance, 
there are some calculations of extra-dimensional
loop effects.
In flat space,
gauge couplings have linear divergence~\cite{%
Dienes:1998vh, 
Dienes:1998vg}.
In warped space, the divergence of
gauge couplings is logarithmic~\cite{%
Pomarol:2000hp, 
Randall:2001gb}.
Taking into account
the form of
the background 
may give a clue to address
the problem of high degrees of divergence
for loop integrals.

Extra dimensions need to be hidden if they exist.
At low energies spacetime is 
effectively in four dimensions
and at high energies extra dimensions are visible
or their signals are visible.
It is possible to describe well this picture 
in a method 
of tracking the number of Kaluza-Klein (KK) 
modes~\cite{%
Bhattacharyya:2006ym, 
Uekusa:2007im}.
In this method, the coupling has linear dependence
on the number of KK modes.
At higher energies the number of KK modes
is proportional to the cutoff so that
dependence is linear to the cutoff.
In a typical theory
with flat extra dimensions, coupling constants
quickly grow and become non-perturbative at about
${\cal O}(10)$
times larger than
the KK 
scale. 
If the higher-dimensional theory
is a perturbation theory 
only right above the KK 
scale,
the region where it is regarded as a
quantum field theory
would be too small.
Flat extra dimensions may not be suitable
for perturbation theory of quantum 
fields in higher dimensions.
In warped space,
a position-dependent cutoff
directly leads to  extra-dimensional signals
without tracking the number of KK modes.
The pioneering idea with warped space
was to generate the hierarchy 
by making
the cutoff on the Planck brane be 
such a large scale as the Planck scale~\cite{%
Randall:1999ee}.
Here we stress it is nontrivial whether the idea
of the hierarchy solution
is compatible with the aspect that
at the Planck scale 
gravity may be quantized.
When the Planck scale is included in the context
of a theory,
the geometry does not appear to be 
fixed in a classical form.
Therefore, 
if there is an extra-dimensional theory as a quantum 
theory without including quantization of 
gravity, 
the candidate might be a theory
in warped space at an intermediate scale 
whose cutoff is much larger than the TeV scale
and much less than the Planck scale.

In this paper, we study
a quantum field theory in warped space
at an intermediate 
scale between the TeV and the Planck scales.
The intermediate scale is regarded as
being much lower than the scale of
the geometry stabilization or the Planck scale.
In order to avoid the appearance
of the Planck scale in the present 
framework, 
we place
the intermediate and TeV branes at the ends
of the bulk
instead of the Planck and TeV branes.
In dealing with
the action integral of field theory,
one of the important things is to have a guide of
invariance of theory.
On the one hand,
Lorentz invariance in the bulk is explicitly broken.
On the other hand, gauge invariance tends to
be preserved at high energies~\cite{Uekusa:2010kh}.
In this light,
we 
deal with gauge theory, although the estimation
of the effect
of the extra-dimensional Lorentz violation 
is beyond the scope of this paper.
The vertices of our interest are
interactions for low-mass states which are
dynamical in effective theory.
We examine
vertex corrections to self-couplings
where zero modes are in external lines.
Introducing a position-dependent cutoff,
we find that it is possible to perform
diagrammatic calculations
where 
the contributions include the effect
of a virtual bulk propagation. 
The 
values of the corrections
are not so different
from the four-dimensional correspondent
and this 
supports validity of perturbation
at the intermediate scale.
On the other hand, it is found that
there is 
nontrivial dependence of the corrections
 on the curvature and the warp factor
and that 
for a part of each diagram 
the coefficient for the logarithmic scaling
has the value in the four-dimensional theory.
In addition to the problem of validity of 
perturbation,
we discuss the problem of higher-dimensional 
operators associated with
non-renormalizability.
Non-renormalizability is linked to
infinite number of counterterms.
We point out a direction to try
to identify vertex corrections for low-mass states 
from higher-dimensional operators
without knowing details about 
how higher-dimensional operators themselves
are generated and receive corrections.

This paper is organized as follows.
In Sec.~\ref{sec:v},
an outline of calculations of vertex corrections
with a position-dependent cutoff is given.
In Sec.~\ref{sec:model}, the model of
non-Abelian theory in warped space is given.
In Sec.~\ref{sec:loop}, our result of loop corrections
is shown.
In Sec.~\ref{sec:high}, higher-dimensional
operators are discussed.
We conclude in Sec.~\ref{sec:conclude}.
Details of Green functions and formula
of Bessel functions
are summarized in Appendices~\ref{app:green}
and \ref{app:bessel}, respectively.

\section{Position-dependence of vertex corrections
\label{sec:v}}

Our interest in this paper is
quantum aspects of
vertices in five-dimensional spacetime.
As 
two external lines correspond to propagators,
the first vertex appears when the number
of external lines is three.
With taking into account a three-point vertex,
we can imagine the further extension
to multiple point diagrams
and higher loop corrections.
The treatment of two-point diagrams
can be read from that of three-point diagrams
so that the comparison of the two-point diagrams
with the three-point diagrams is applied to
the analysis of multiple point diagrams.
In addition, in the four-dimensional theory
the two-loop renormalization for two-point diagrams
needs the one-loop counterterms for three-point 
diagrams.

\begin{figure}[htb]
\begin{center}
\includegraphics[width=3cm]{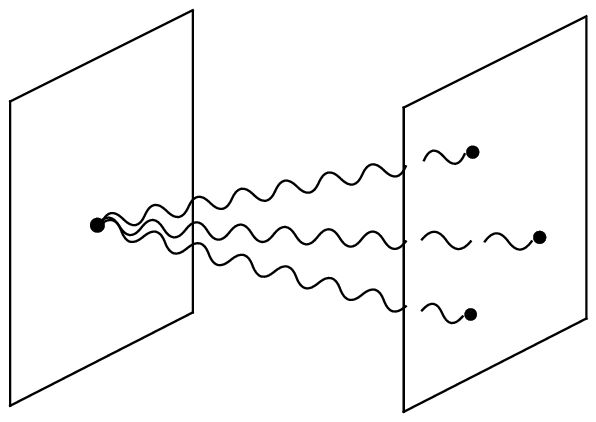}
\qquad
\includegraphics[width=1cm]{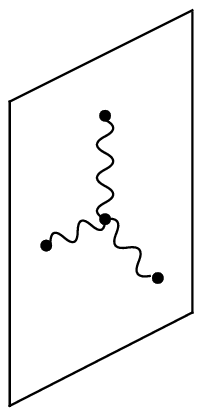}
\end{center}
\caption{Tree-level vertices. \label{fig:vtree}}
\end{figure}

\begin{figure}[htb]
\begin{center}
\includegraphics[width=3cm]{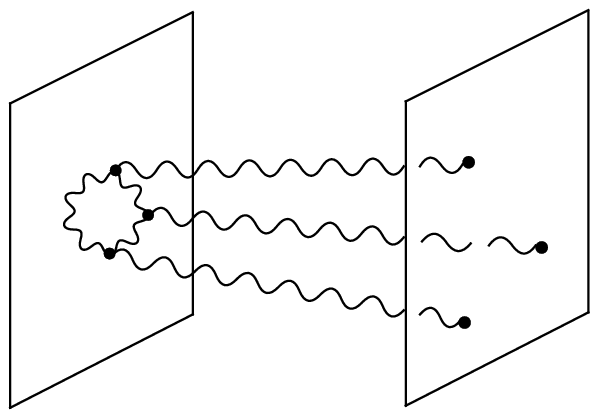}
\qquad
\includegraphics[width=1cm]{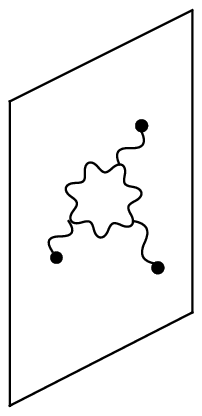}
\end{center}
\caption{One-loop vertices. \label{fig:vloop}}
\end{figure}

For the external lines with fixed positions,
their vertices are located at various positions in 
five-dimensional spacetime.
In 
Figure~\ref{fig:vtree}, four-dimensional
directions are denoted in vertical directions
and the extra-dimensional direction is denoted in
a horizontal direction. 
Each vertex has corrections. 
As a practical way to deal with warped space,
we adopt a position-dependent cutoff for
four-momentum.
This is important for keeping perturbative validity
of coupling constants.
Examples of one-loop diagrams are given
in Figure~\ref{fig:vloop}.
For the left and right figures for vertices in
Figure~\ref{fig:vloop},
the momentum
cutoffs of loop integral 
are taken differently from
each other.
All of these position-dependent contributions must be
summed up continuously, i.e., must be integrated.

\begin{figure}[htb]
\begin{center}
\includegraphics[width=6cm]{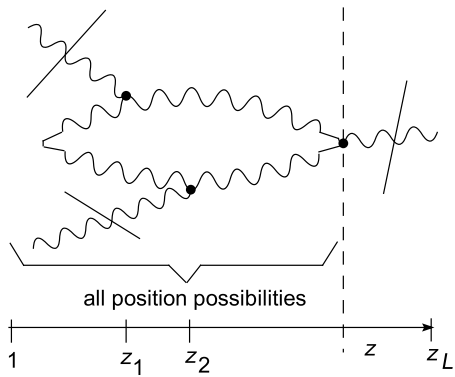}
\caption{Vertex correction for zero mode  
\label{vzero} }
\end{center}
\end{figure}

\begin{figure}[htb]
\begin{center}
\includegraphics[width=6cm]{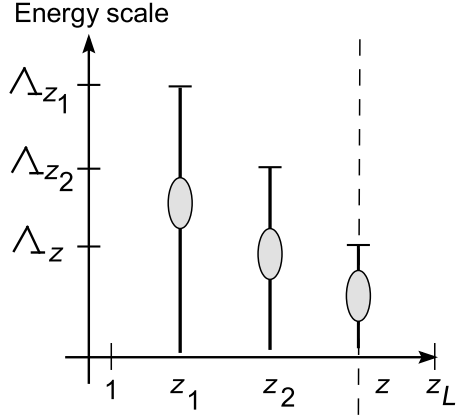}
\caption{The energy region of the theory
depending on $z$. 
The typical energy scale of the theory
at a position $z$ is denoted as a circle.
\label{fig:nrg}}
\end{center}
\end{figure}

It 
is physically important to consider
the situation that
dynamical fields in effective theory below
the TeV scale
are external lines for three-point functions.
For pure gauge theory without symmetry breaking,
such dynamical fields are massless mode.
The zero mode is a constant with respect to
the position in the extra dimension.
We will explicitly show in the next section
 that the Green function 
connecting a five-dimensional field to 
a four-dimensional zero mode is position-independent. 
Because of this position independence,
loop corrections can be calculated for
diagrams where 
external lines are amputated.
For example, a one-loop diagram composed of
three-point interactions is shown in Figure~\ref{vzero}.
The cutoff for the four-momentum is determined
by the 
largest
interacting position $z$ shown as 
the vertical broken line in Figure~\ref{vzero}.
This diagram involves the loop- and $z$-integrals
given by
\bea
    \int_1^{z_L} {dz\over kz}
      \int_{|\ell|\leq \Lambda_z} 
       {d^4 \ell \over (2\pi)^4}
     \int_1^{z} {dz_1 \over kz_1}
     \int_1^{z} {dz_2 \over kz_2}
       G_\ell (z,z_1) 
        G_\ell (z, z_2) G_\ell( z_1, z_2) \times F,
    \label{maineq}    
\eea
where a Green function 
is denoted as $G_p(z,z')$
and it depends on
four-momentum and the fifth-dimensional position.
The factor $F$ is associated with momentum flow
and it generally includes the loop momentum $\ell$.
The position-dependent cutoff has been introduced
as $\Lambda_z \equiv \Lambda z_L/z$.
The momentum cutoff at $z=z_L$ is $\Lambda$
and is assumed as $\Lambda \simeq 10$TeV.
For a small $z$, $\Lambda_z$ becomes large.
The energy region of the theory depending on $z$
is schematically shown in Figure~\ref{fig:nrg}.
The loop four-momenta run up to the common energy
scale, $\Lambda_z$ in Figure~\ref{fig:nrg}.
The cutoff of the theory at a position
and also the common energy scale
do not need to reach
the typical energy region
of the theory at a different position,
as seen for the theories at $z_1$ and $z$
in Figure~\ref{fig:nrg}.
The other diagrams will be taken into account similarly.

\section{Non-Abelian gauge theory in warped space
\label{sec:model}}

In this section, we give the model
and the resulting two-point functions and Green 
functions.
The action integral for non-Abelian gauge theory
is given by
\bea
  && \int d^4 x \int_1^{z_L} 
  d z \sqrt{\textrm{det}(g_{K L})}
    \textrm{tr}
      \left( 
      -{1\over 2} F_{MN} F_{P Q} g^{M P} g^{N Q}
    -{1\over \xi}
    \omega(A)^2 \right) 
  + S_{\textrm{\scriptsize ghost}} ,
   \label{ai}
\eea
with the metric in the five-dimensional
spacetime
\bea
   ds^2 =
      {1\over z^2}
        (\eta_{\mu\nu} dx^\mu dx^\nu
           -{1\over k^2} dz^2) .
\eea
The four-dimensional Minkowski metric is
$(+1, -1, -1, -1)$.
The extra-dimensional coordinate is denoted as
$z$. The coordinate $z$ takes a value
in $1\leq z \leq z_L \equiv e^{kL}$,
where $k$ is the curvature and $L$ is 
the size of the extra dimension.
The five-dimensional gauge field $A_M$
is parted into $A_\mu$ and $A_z$.
The gauge fixing can be taken so as to
remove kinetic mixing terms,
\bea
\omega(A) =
\partial_{\mu} A_{\nu} \cdot
     g^{\mu\nu}
     + \xi z g^{zz} \partial_z
       ({1\over z} A_z) .
\eea
Correspondingly to this gauge fixing,
the ghost action integral is given by
\bea
    S_{\textrm{\scriptsize ghost}}
   &\!\!\!=\!\!\!&
   \int d^4 x \int_1^{z_L}
   {dz\over kz}
  \left( \bar{c}^a (-\partial^\mu \partial_\mu) c^a
   -g f^{abc}  \bar{c}^a \partial^\mu
  (A_\mu^b c^c) 
  \right.
\nonumber
\\
  && \left.
   +\xi \bar{c}^a k^2 z \partial_z
   ({1\over z} \partial_z) c^a
  +\xi g f^{abc}  
   \bar{c}^a k^2 z \partial_z ({1\over z} A_z^b c^c)
  \right) .
\eea
Here the contraction of a subscript 
with a superscript stands for
a contraction with $\eta^{\mu\nu}$ such as
$F_{\mu \nu} F^{\mu \nu} 
=F_{\mu\nu} F_{\rho\sigma} \eta^{\mu\rho}
\eta^{\nu\sigma}$.
A contraction with $g^{M N}$ is described
explicitly with $g^{MN}$ such as 
$F_{\mu \nu} F_{\rho \sigma} g^{\mu\rho} g^{\nu\sigma}$.
This rule will be used throughout 
this paper.
From the action integral (\ref{ai}),
interactions for $A_\mu$ and $A_z$ are given by
\bea
   &&
     \int d^4 x \int_1^{z_L} {dz\over kz}
       \left( -g f^{abc}
         \partial_\mu A_\nu^a \cdot
           A^{\mu b} A^{\nu c}
    -{1\over 4} g^2 f^{ab e} f^{c de}
       A_\mu^a A_\nu^b A^{\mu c} A^{\nu d}
       \right.
\nonumber
\\
  && \left.
    +k^2 g f^{abc} 
      (\partial_\mu A_z^a
        -\partial_z A_\mu^a)
         \cdot A^{\mu b} A_z^c
    +{1\over 2} k^2 g^2 
      f^{ab e} f^{c de} A_\mu^a A_z^b
         A^{\mu c} A_z^d \right) .
         \label{int}
\eea
In the equation (\ref{int}),
the first term with three gauge fields being zero mode
is the vertex of our interest.
We will examine how the term receives corrections.

In order that 
symmetry group before the gauge fixing
in four dimensions is identical with
the original gauge group,
$A_\mu$ and ghost fields
obeys Neumann boundary condition and
$A_z$ obeys Dirichlet boundary condition.
From equations of motion and boundary conditions,
the zero modes for $A_\mu$ and ghost are constants
with respect to the extra-dimensional coordinate.
The gauge field is expanded as
\bea
   A_\mu (x,z ) =
     {1\over \sqrt{L}}
       A_{\mu 0} (x) 
     + \sum_{n=1}^\infty
        N_n A_{\mu n}(x) \chi_n (z) ,
\eea
where $N_n$ is a normalization constant.
For the $\xi =1$ gauge,
massive-mode function $\chi_n$ for $A_\mu$ 
is given by
\bea
   \chi_n (z)
  =z (J_1({m_n z\over k})  +\beta_n Y_1
  ({m_n z\over k}) ) ,
\eea
where
$\beta_n
=-J_0(m_n/k)/
  Y_0 (m_n/k)
=-J_0(m_n z_L/k)/
  Y_0(m_n z_L/k)$.
The mode function $\phi_n$ for $A_z$ is given by
$\phi_n(z) = (k/m_n) \partial_z \chi_n (z)$.

The Green functions obeying the following 
equations can be introduced,
\bea
   &&
   \left(\partial_z^2 -{1\over z} \partial_z
     +{p^2 \over k^2} \right)
       G_p (z,z') 
        = {z\over k} \delta(z-z') ,
\\
  &&
    \left( \partial_z^2 -{1\over z}\partial_z
      +{p^2 \over k^2 \xi} \right)
        G_{p\over \sqrt{\xi}}
          (z,z') 
          = {z\over k} \delta(z-z') ,
\\
 && \left(
    \partial_z^2 -{1\over z} \partial_z
      +{1\over z^2}
        +{p^2 \over k^2 \xi}\right)
        S_{p\over \sqrt{\xi}} (z,z') =
        {z\over k} \delta(z-z') .
\eea
With these Green functions,
two-point functions in the $\xi=1$ gauge for
$A_\mu$, $A_z$ and ghost are given by
\bea
  \langle
  \tilde{A}^{\mu a} (p,z')
    \tilde{A}^{\nu b} 
   (q, z) \rangle
   &\!\!\!=\!\!\!&
   \delta^{ab} (-i G_p (z,z') \eta^{\mu\nu}) 
  (2\pi)^4 \delta^4 (p+q),
\\
  \langle \tilde{A}_z^a (p,z')
   \tilde{A}_z^b (q, z)\rangle
  &\!\!\!=\!\!\!&
    \delta^{ab} {1\over k^2} i S_p (z,z') 
   (2\pi)^4 \delta^4 (p+q),  
\\
  \langle \bar{\tilde{c}}^a
    (p,z') \tilde{c}^a (q, z) \rangle
      &\!\!\!=\!\!\!& i G_p (z,z')
   (2\pi)^4 \delta^4 (p+q) ,
\eea
where the four-dimensional coordinates
are written in a momentum picture
via Fourier transformation. 
The explicit forms of Green functions are shown
in Appendix~\ref{app:green}.

The two-point function for a five-dimensional
gauge field and a four-dimensional zero-mode gauge field
is found from the orthogonality of the mode function.
The orthogonality of $\chi_n$
yields
\bea
  \int_1^{z_L}
    {dz \over kz} A_\mu (x,z) {1\over \sqrt{L}}
       =A_{\mu 0}(x) .
\eea
With this property and
the integral of the Green function
\bea
  && \int_1^{z_L}  {dz' \over kz'}
     G_p  (z, z') 
 =
    \int_1^z {dz'\over kz'}
      G_{p} (z,z')_{z'\leq z}
       +\int_z^{z_L} {dz'\over kz'}
        G_{p} (z,z')_{z'\geq z}
  =
  {1\over p^2 } ,
    \label{gchi}
\eea
the
two-point function for $A_\mu$ and $A_{\nu 0}$ is
obtained as
\bea
   \langle \tilde{A}_\mu^a (p,z)
    \tilde{A}_{\nu 0}^b (q) \rangle
  ={1\over \sqrt{L}}
    {1\over p^2}
      \delta^{a b} (2\pi)^4 
        \delta^4 (p+q) (-i\eta_{\mu\nu}) .
   \label{aa0}
\eea
The two-point function (\ref{aa0})
is independent of the position
$z$.

For KK mode, the mode function has
the $z$-dependence.
In analogy with Eq.~(\ref{gchi}),
the formula is given by
\bea
    \int_1^{z_L}
    {dz' \over kz'} 
      G_p (z,z') \chi_n (z')
      ={1\over p^2 -m^2} \, \chi_n(z) .
\eea
From this equation, 
the two-point function for a 
five-dimensional gauge field and a four-dimensional
KK-mode gauge field is obtained as
\bea
   \langle 
     \tilde{A}_\mu^a (p,z)
       \tilde{A}_{\nu n}^b (q) \rangle
     =N_n \chi_n(z) {1\over p^2 -m^2}
     \delta^{ab} (2\pi)^4 
      \delta^4(p+q) (-i \eta_{\mu\nu}) .
  \label{5andkk}
\eea 
Diagrams with KK-mode external lines 
may be amputated with respect to
the four-dimensional part,
whereas the $z$-dependence such as
$N_n \chi_n(z)$ in Eq.~(\ref{5andkk}) needs to be
taken into account in Eq.~(\ref{maineq}).

\section{Loop corrections \label{sec:loop}}

We calculate
loop corrections of the three-point vertex
in the model given in the previous section.
For the one-loop diagrams, the tensor structure
for momentum flow is analogous to the four-dimensional
case.
The Green function for gauge field is related to
the four-dimensional gauge field propagator
with the correspondence,
\bea
   -{i \eta_{\mu\nu} \over p^2 }
  \leftrightarrow i G_p (z,z') \eta_{\mu\nu} ,
\eea
in the $\xi=1$ gauge.
Using the same Feynman parameters as in
four-dimensional case,
we find the one-loop divergent part 
\bea
 {\cal M}_{\textrm{\scriptsize 1-loop},
   \mu\nu\rho}^{abc} =  
   {1\over (4\pi)^2} 
{g^3\over L\sqrt{L}}
      C_2 (G) f^{abc}
   V_{\mu\nu\rho}  
  {\cal M}_{\textrm{\scriptsize 1-loop}},
    \label{onediv}
\eea
with 
\bea
   {\cal M}_{\textrm{\scriptsize 1-loop}}
 = -2 \left[
   \left({13\over 8} -{1\over 24}\right) 
     {U_1 \over 2}     
   - {9\over 4} U_2 \right] ,
   \label{bloop}
\eea  
where the factors $13/8$, $-1/24$
and $-9/4$ arise
from three-point interactions,
a ghost loop and
the contribution with a four-point interaction,
respectively.
The tensor structure
for the momentum flow of the external lines is 
given by
\bea
V_{\mu\nu\rho} =
          (q+2 p)_\nu \eta_{\mu \rho}
          -(p+2 q)_\mu \eta_{\nu \rho}
          +(q-p)_\rho \eta_{\mu \nu} .
\eea  
The loop- and $z$-integrals are included in 
\bea
  i{2\over (4\pi)^2} 
  U_1 &\!\!\!=\!\!\!&  
 \int_1^{z_L} {dz\over kz}         
   \int_{|\ell|\leq  \Lambda_z} 
      {d^4 \ell\over (2\pi)^4}
        \int_1^{z} 
          {d z_1\over k z_1}  
        \int_1^{z} 
           {d z_2 \over k z_2}
   G_\ell (z, z_1) G_\ell (z, z_2) G_\ell (z_1, z_2)
     \ell^2 ,
\\
  i {2\over (4\pi)^2}
  U_2 &\!\!\!=\!\!\!&
    \int_1^{z_L} {dz\over kz}
  \int_{|\ell|\leq \Lambda_z}
   {d^4 \ell\over (2\pi)^4}
  \int_1^{z}
   {d z_1 \over k z_1}     
   (G_\ell (z, z_1))^2 .
\eea
These needs to be examined with a numerical analysis.
Then ${\cal M}_{\textrm{\scriptsize 1-loop}}$
is obtained.
The correspondence of $U_1$ and $U_2$ with
the four-dimensional case is 
\bea
   U_1 &\!\!\! \leftrightarrow \!\!\!&
   \left[ i {2\over (4\pi)^2} \right]^{-1}
     \int {d^4 \ell \over (2\pi)^4}
  {2 \ell^2 \over \left[\ell^2 -\Delta\right]^3}
    = 
 \log \left(\Lambda^2 \over 
  \Delta\right) ,
    \label{u1}
\\  
  U_2 &\!\!\! \leftrightarrow \!\!\!&
      \left[ i {2\over (4\pi)^2} \right]^{-1}
  \int {d^4 \ell \over (2\pi)^4}
  {1 \over \left[\ell^2 -\Delta\right]^2}
    = {1\over 2} \log \left(\Lambda^2 \over 
  \Delta\right) ,
    \label{u2}
\eea
where $\sqrt{|\Delta|}$ has 
the scale of external momenta. 
The four-dimensional
correspondent for Eq.~(\ref{bloop})  
is given by
\bea
  {\cal M}_{\textrm{\scriptsize 1-loop}} 
   \leftrightarrow
    {2\over 3} \log 
  \left({\Lambda^2 \over \Delta}\right)
    \simeq {4\over 3} \log \left({10 \textrm{TeV}
          \over 100 \textrm{GeV}}\right)
      = {8\over 3} \log 10 .
      \label{4d}
\eea 
Since the momentum-dependence of Eq.~(\ref{4d})
is logarithmic,
the order of the value is not sensitive to
a choice of the value of $\Delta$.

\begin{figure}[htb]
\begin{center}
\includegraphics[width=7cm]{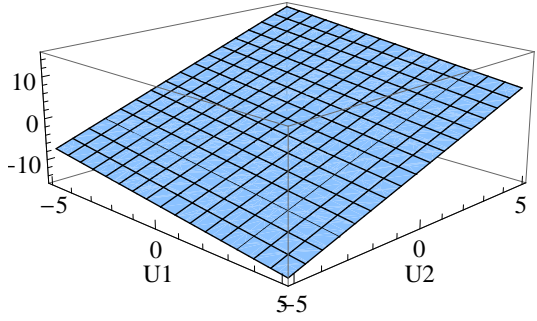}
\caption{The dependence of 
$\overline{\cal M}_{\textrm{\scriptsize 1-loop}}$ on
$\bar{U}_1$ and $\bar{U}_2$.
\label{fig:b}}
\end{center}
\end{figure}

Now we analyze $U_1$ given in Eq.~(\ref{u1}).
After the $z_1$ and $z_2$ integrals and the 
Wick rotation, 
$U_1$ is written as
\bea
   U_1 &\!\!\!=\!\!\!& 
 - \int_1^{z_L} {dz\over kz}         
   \int_0^{\Lambda_z} 
      d \ell_E
       I_{G 3} (z)
     \ell_E^5 ,
\eea
where the integral of the product
of the Green functions is given by
\bea
  I_{G 3} (z)
     &\!\!\!=\!\!\!&
   {1\over k^5} \bigg\{
     - {\bar{N}^2 \over 8}
          (\bar{d}_1(z))^2  {1\over P_E^2}
            \left[
            {4\bar{N} \over \pi^2} T(1)
              + 2 a_0 (z) a_2(z)
               \right.
\nonumber
\\
  && \left.
    - \bar{N} P_E^2 T(z)
      \left[ (\bar{a}_1(z))^2
     + a_0(z) a_1(z) \right] \right]
         \bigg\} .
         \label{IG3con}
\eea
The functions appearing in Eq.~(\ref{IG3con}) are
written in terms of the modified Bessel functions
as
\bea
   \bar{N} &\!\!\!=\!\!\!&
  {\pi^2 \over 4} 
  {1\over K_0(P_E) I_0 (P_E z_L)
   - I_0 (P_E) K_0 (P_E z_L) } ,
\\
    \bar{d}_1 (z)
  &\!\!\!=\!\!\!&
   {2\over \pi} z
  \left[ K_0 (P_E z_L) I_1 (P_E z)
   + I_0 (P_E z_L) K_1 (P_E z) \right] ,
\\
  T(z) &\!\!\!=\!\!\!&
   -2 \bar{a}_1 (z) \bar{d}_1 (z)
  -a_0 (z) d_2 (z) -a_2 (z) d_0 (z) ,
\\
  d_0 (z) &\!\!\!=\!\!\!&
   {2\over \pi} z
  \left[ K_0 (P_E z_L) I_0 (P_E z)
     - I_0 (P_E z_L) K_0 (P_E z) \right] ,
\\
  d_2 (z) &\!\!\!=\!\!\!&
   {2 \over P_E z} \bar{d}_1 (z) - d_0 (z) ,
\eea
where $\ell_E$ denotes
the length in the polar coordinate
and $P_E = \ell_E/k$. 
The functions $a_i (z)$ are given by
$d_i (z)$ with the replacement $z_L \to 1$.        
The detail of the derivation of Eq.~(\ref{IG3con})
is shown in 
Appendix~\ref{app:green}.
Similarly $U_2$ is obtained as
\bea
   U_2 =  
     \int_1^{z_L}  {dz\over kz}
     \int_0^{\Lambda_z}
    d\ell_E I_{G 2} (z,z) 
     \ell_E^3 ,
\eea
where 
\bea
   I_{G 2} (z,z)
    = {1\over k^3}
       \left\{
        {1\over 2}
          \bar{N}^2 (\bar{d}_1(z))^2
           \left[
           (\bar{a}_1 (z) )^2
            +a_0(z) a_2(z)
  - {4 \over \pi^2 P_E^2} \right] 
 \right\} .
\eea
We evaluate $\bar{U}_1 =U_1/\log z_L$ 
and $\bar{U}_2 =U_2/\log z_L$ changing
the values of $k$ and $z_L$.
These values are order ${\cal O}(1)$
at scales much larger than
the TeV scale such as
$\Lambda_z|_{z=1} =10^{11}\textrm{GeV}$
for $z_L =10^7$.
The numerical values of $\bar{U}_j$, $j=1,2$
are tabulated in Table~\ref{tab:u}.
\begin{table}[htb]
\begin{center}
\caption{$\bar{U}_1 = U_1/\log z_L$ and 
$\bar{U}_2 =U_2/\log z_L$. 
For comparison, 
the four-dimensional correspondents are denoted.
\label{tab:u}}
\vspace{0.4ex}
\begin{tabular}{c|c|c|c|c|c|c|c|c|c||c}
\hline\hline
$k/z_L$
& \multicolumn{3}{c|}{$10^3$GeV}
& \multicolumn{3}{c}{$10^5$GeV}
& \multicolumn{3}{|c||}{$10^7$GeV} 
& 4D theory 
\\
\hline
$z_L$ & $10^{11}$ & $10^7$ & $10^3$  
& $10^{11}$ & $10^7$ & $10^3$  
& $10^{11}$ & $10^7$ & $10^3$  
& $\Lambda/\Delta^{1/2}$
\\
%
\hline
$\bar{U}_1$ &
0.88 & 1.02 & 1.34 
& 0.44 & 0.48 & 0.75
& 0.40 & 0.42 & 0.73 
& 1 
\\
$\bar{U}_2$ &
1.51 & 1.52 & 2.12 
& 0.55 & 0.70 & 1.08 
& 0.52 & 0.62 & 1.06
& 0.5 
\\
\hline
\end{tabular}
\end{center}
\end{table}
From Eqs.~(\ref{u1}) and (\ref{u2}),
the four-dimensional correspondents 
satisfy
$U_1=2U_2$.
The contributions in the warped space
tend to have
\bea
   U_1  = {1\over 2} \bar{U}_1 \log z_L^2
 < 2 U_2 =\bar{U}_2 \log z_L^2 , 
 \label{u12u2}
\eea
The correspondence between the values 
in the four-dimensional
case and the warped case occurs depending
on $k$ and $z_L$.
The value 
$\bar{U}_1 \approx 1$ is generated
for a relatively small $k/z_L$.
The value $\bar{U}_2 \approx 0.5$ is generated
for a large $z_L$ and a relatively large $k/z_L$.
Once the factor $U_1$ and $U_2$ are obtained,
the contribution 
${\cal M}_{\textrm{\scriptsize 1-loop}}$
is derived from Eq.~(\ref{bloop}).
The behavior of 
$\overline{\cal M}_{\textrm{\scriptsize 1-loop}}
={\cal M}_{\textrm{\scriptsize 1-loop}}/\log z_L^2$ 
as a function of $\bar{U_j}$ 
is shown in Figure~\ref{fig:b}.
It is seen that
the value 
$\overline{\cal M}_{\textrm{\scriptsize 1-loop}}$ 
can be positive or negative
in a region of $\bar{U}_j \sim {\cal O}(1)$.   
The numerical values of 
$\overline{\cal M}_{\textrm{\scriptsize 1-loop}}$
for the parameters $z_L$ and $k$ 
given in Table~\ref{tab:u}
are tabulated in Table~\ref{tab:b}.
\begin{table}[htb]
\begin{center}
\caption{$\overline{\cal M}_{\textrm{
\scriptsize 1-loop}}
={\cal M}_{\textrm{\scriptsize 1-loop}}/
\log z_L^2$. 
\label{tab:b}}
\begin{tabular}{c|c|c|c|c|c|c|c|c|c||c}
\hline\hline
$k/z_L$
& \multicolumn{3}{c|}{$10^3$GeV}
& \multicolumn{3}{c}{$10^5$GeV}
& \multicolumn{3}{|c||}{$10^7$GeV}
& 4D theory \\
\hline
$z_L$ & $10^{11}$ & $10^7$ & $10^3$  
& $10^{11}$ & $10^7$ & $10^3$  
& $10^{11}$ & $10^7$ & $10^3$
& $\Lambda/\Delta^{1/2}$  \\
\hline
$\overline{\cal M}_{\textrm{\scriptsize 1-loop}}$ &
$2.70$ 
& $3.06$ 
& 3.71
& 0.89 
& 1.20
& 1.84 
& 0.85 
& 1.06 
& 1.81 
& 0.67
\\
\hline
\end{tabular}
\end{center}
\end{table}
The value of $\overline{\cal M}_{\textrm{
\scriptsize 1-loop}}$ is order ${\cal O}(1)$ and
the explicit dependence of the correction
${\cal M}_{\textrm{
\scriptsize 1-loop}}$
on the warp factor is logarithmic, $\log z_L^2$.
In this sense,
the perturbation including bulk contributions
is valid even at scales which 
are the warp factor times larger than the TeV scale.
The value $\overline{\cal M}_{\textrm{\scriptsize
1-loop}} \approx 2/3$ is obtained
for a large $z_L$ and a relatively large $k/z_L$
where $U_2$ in Eq.~(\ref{bloop}) is close 
to the four-dimensional value
and $U_1$ is relatively small.
The feature of the result is summarized as follows.
On the one hand, 
it is found that the 
value of
the factor $\overline{\cal M}_{\textrm{
\scriptsize 1-loop}}$ in Table~\ref{tab:b} 
is not so different from
$2/3$ in the four-dimensional 
case irrespective of the relation (\ref{u12u2}).
On the other hand,
each diagrams to compose the total contribution
can give the same value as in the four-dimensional 
case.

\section{A viewpoint for higher-dimensional operators
\label{sec:high}}

Since the gauge coupling is negative mass dimension
$[g] = [\textrm{mass}]^{-1/2}$,
gauge-invariant higher-dimensional operators
can be infinitely written down.
This effect would need to be taken into account
and the action integral is generally written as
\bea
    \int d^4x \int dz
      \sum_{d=0}^\infty \sum_i
         c_{d,i} g^{2 d} {\cal O}_{d+5,i} ,
\eea
where ${\cal O}_{D,i}$ denote a operator with
dimension $[\textrm{mass}]^D$ for species $i$ and
$c_{d,i}$ is the corresponding coefficient.
However, it is unclear whether
all $c_{d,i}$ can be fixed
in the present framework.
We would like to achieve any
approach to identify corrections to 
low-energy interactions
with leaving part of $c_{d,i}$ unknown.
We have examined corrections for the zero-mode
interaction
in the mode expansion of
the three-point interaction in 
Eq.~(\ref{int}),
\bea
     \int d^4 x 
       \left( -{g \over \sqrt{L}}
   f^{abc}
         \partial_\mu A_{\nu 0}^a \cdot
           A_0^{\mu b} A_0^{\nu c} \right) .
           \label{03}
\eea       
If higher-dimensional operators exist,
they contribute to this vertex.
For example, there would be
dimension-seven operators
\bea
    c_{2,1}  D^P D_P F_{MN} \cdot F^{MN} 
    +c_{2,2} D^M D_P F_{MN} \cdot F^{P N} 
    + \cdots.
\eea
For simplicity, we assume that $c_{2,1}$ is dominant
over the other coefficients.
Then propagator has an extra factor $1/(c_{2,1} p^2)$
for a large momentum.
In addition, three-point and four-point vertices 
have an extra factor $c_{2,1} p^2$ for 
a large momentum.
Because the propagator is suppressed by
an extra factor, the divergent
correction to Eq.~(\ref{03}) arises
not from the original dimension-five operators
but from the dimension-seven operators.
One-loop diagrams are the same form as drawn by
the original dimension-five operators.
The diagram composed of only three-point vertex 
has three internal propagators and three vertices.
The ghost loop is similar.
The diagram including four-point vertex
has two internal propagators and two vertices.
Therefore, the extra factors in the propagator
and the vertices are canceled each other
in the diagrams.
\begin{figure}[htb]
\vspace{2ex}
\begin{center}
\includegraphics[width=9cm]{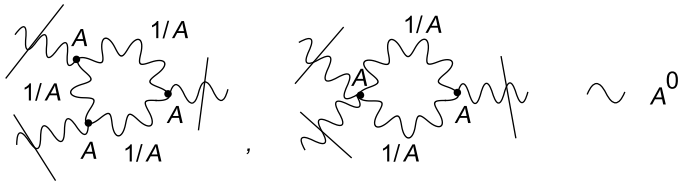}
\end{center}
\caption{Cancellation of
the extra factor $A\equiv c_{2,1}p^2$. 
\label{fig:extraf}}
\end{figure}
This
is depicted in Figure~\ref{fig:extraf}.
Then we do not need to know
what the value of the coefficient $c_{2,1}$ is.
Although the idea here is not a solution of 
the problem of higher-dimensional operators,
further development might lead to
a method to bypath
the problem of higher-dimensional operators.

\section{Conclusion \label{sec:conclude}}

We have studied
a quantum field theory in warped space
at an intermediate 
scale between the TeV and the Planck scales.
Our point has been 
to avoid the appearance
of the Planck scale in the present 
framework by placing
the intermediate and TeV branes at the ends
of the bulk 
instead of the Planck and TeV branes.
For this setup, we have examined
vertex corrections to self-coupling
where zero modes are in external lines.
In the present case, zero mode is dynamical 
below the TeV scale.
The effect of a position-dependent cutoff
has been included via the extra-dimensional integral.
The cutoff for the four-momentum 
$\Lambda_z = \Lambda z_L /z$ is determined
by the smallest interacting point $z$.
It has been shown that there is also
a technical advantage
with zero mode in external lines. 
The Green function 
connecting a five-dimensional field to 
a four-dimensional zero mode is position-independent
so that calculations are performed
for amputated diagrams.
After integrals of products of
the five-dimensional Green functions,
we have found 
the vertex correction ${\cal M}_{\textrm{
\scriptsize 1-loop}}$
which depends
on the curvature and the warp factor.
The values of the corrections
are not so different
from the four-dimensional correspondent.
Its explicit dependence
on the warp factor is logarithmic, $\log z_L^2$.
This supports validity of perturbation
at the intermediate scale.
As for the correspondence with the four-dimensional
theory, the value 
$\bar{U}_1 \approx 1$ is generated
for a relatively small $k/z_L$.
The values $\bar{U}_2 \approx 0.5$ 
and $\overline{M}_{\textrm{\scriptsize 1-loop}}
\approx 2/3$
are generated
for a large $z_L$ and a relatively large $k/z_L$.
In addition to the issue of 
perturbation,
we have discussed the problem of higher-dimensional 
operators associated with
non-renormalizability.
We have pointed out
the possibility that the extra factor
originated from a higher-dimensional operator
is canceled in the loop diagrams
for the zero-mode self-coupling vertex.
It still remains to examine 
higher-dimensional aspects from various viewpoints.
It needs to be investigated further
how a higher-dimensional theory can be
a quantum field theory.

\vspace{8ex}

\subsubsection*{Acknowledgments}

This work is supported by Scientific Grants 
from the Ministry of Education
and Science, Grant No.~20244028.

\newpage

\begin{appendix}

\section{Green functions \label{app:green}}

\subsection{Solutions}

For the variables 
$u\equiv \textrm{min} (z,z')$, 
$v\equiv \textrm{max} (z,z')$,
and
$P \equiv \sqrt{p^2 /k^2}$,
the Green functions are given by
\bea
  G_p (u,v) 
    = N a_1 (u) d_1(v) , \qquad
  S_p (u,v) 
    = N a_0 (u) d_0(v)   
\eea
Here the $u$- and $v$-dependent parts are 
\bea
   a_i(u) =
   u ( A J_i (Pu) + BY_i (Pu) ) ,
\qquad
    d_i (v) =
    v ( C J_i (P v) + D Y_i (P v)) .
\eea
and the constants are given by
\bea   
   A &\!\!\!=\!\!\!& -Y_0 (P) , \qquad
   C = - Y_0 (P z_L) , \qquad
   N = {1\over 2 k} {\pi \over AD -BC} ,
\\ 
   B &\!\!\!=\!\!\!& J_0 (P) , \qquad
    D = J_0(P z_L) .
\eea
The function $G_p$ satisfies Neumann condition
\bea
   \partial_u G_p (u,v) \bigg|_{u=1}
     = \partial_v G_p (u,v) \bigg|_{v=z_L} = 0.
\eea
The function $S_p$ satisfies Dirichlet condition
\bea
    S_p (u,v) \bigg|_{u=1}
     = S_p (u,v) \bigg|_{v=z_L} = 0.
\eea

\subsection{Integrals of products of Green functions}

One of general integrals of two Green functions 
is 
\bea 
  I_{G 2} (z, z_1)&\!\!\!=\!\!\!& 
    \int_1^{z}
         {dz_2\over kz_2} 
           G_\ell (z_1, z_2)
            G_\ell (z,z_2)
\nonumber
\\        
 &\!\!\!=\!\!\!&
    \int_1^{z_1}
         {dz_2\over kz_2} 
           G_\ell (z_1, z_2)_{z_2\leq z_1}
            G_\ell (z,z_2)_{z_2\leq z}
        +\int_{z_1}^{z}
           {dz_2 \over kz_2}
             G_\ell (z_1, z_2)_{z_2\geq z_1}
              G_\ell (z,z_2)_{z_2\leq z} 
\nonumber
\\
  &\!\!\!=\!\!\!&
     {N^2 \over k} d_1(z_1) d_1(z) 
       \int_{1}^{z_1}
         {dz_2 \over z_2} (a_1(z_2))^2 
     + {N^2 \over k} a_1(z_1) d_1(z)
        \int_{z_1}^{z}
          {dz_2\over z_2}  d_1(z_2) a_1(z_2) ,
  \label{IG2}
\eea
where $P = \sqrt{\ell^2/k^2}$ has been substituted.
The integrals for the above two terms are given by
\bea
      \int^z
        {dz' \over z'} (a_1(z'))^2
  &\!\!\!=\!\!\!&
        {1 \over 2}\left\{
          (a_1(z))^2
     -a_0 (z) a_2 (z) \right\} ,
\\
      \int^z {dz' \over z'}
         a_1 (z' ) d_1(z')
  &\!\!\!=\!\!\!&
        {1\over 4} T(z) ,
\eea
with 
$T(z) \equiv 2  a_1 (z) d_1 (z) 
      -a_0 (z)  d_2 (z)
    -a_2 (z) d_0 (z)$.
The equation (\ref{IG2}) is written as
\bea
    I_{G 2} (z,z_1)
    &\!\!\!=\!\!\!&
       F_1 (z) d_1 (z_1) 
     + F_2 (z)  z_1 a_0 (z_1) 
     + F_3 (z) a_1 (z_1) .
\eea
Here 
\bea
  F_1 (z) &\!\!\!= \!\!\!&
    {N^2 \over k} {2\over \pi^2 P^2} d_1(z) ,
\qquad
  F_2 (z) = 
    - {N \over 2 k^2 P} d_1(z) ,
\\
  F_3 (z) &\!\!\!= \!\!\!&
       {N^2 \over 4 k} d_1(z)  
  \left\{
     {2\over N k P^2} + T(z)
      \right\} .
\eea

The integral of three Green functions appearing in
one-loop diagrams is 
\bea
  I_{G 3} (z)&\!\!\!=\!\!\!&
     \int_1^{z}
       {dz_1 \over kz_1}
       \int_{1}^{z}
         {dz_2\over kz_2} 
        G_\ell (z ,z_1)  
           G_\ell (z_1, z_2)
            G_\ell (z,z_2)
\nonumber
\\
  &\!\!\!=\!\!\!&
   {N\over k} d_1(z) F_1 
   \left[ {1\over 4} 
  T(z_1) \right]_1^{z}
  +{N\over k} d_1(z) F_2 
   \left[ {1\over 2 P}
   (a_1 (z_1) )^2 \right]_1^{z} 
\nonumber
\\
  &&
    +{N\over k} d_1(z) F_3 
    \left[
      {1\over 2} \left\{
         (a_1 (z_1))^2
            - a_0 (z_1) a_2 (z_1) \right\}
              \right]_1^{z}
   . \label{IG3}
%
\eea
Here the formula 
\bea
   \int^z dz'
       (a_1 (z') a_0 (z'))
 =
   {1\over 2 P} (a_1 (z_1))^2  ,
\eea
has been used.
The equation ~(\ref{IG3}) can be arranged further
via the Wick rotation.
For $P = i P_E$, 
the functions $a_i$ and $d_i$ are written as
\bea
   a_1 (z)
     &\!\!\!=\!\!\!&
        {2 i \over \pi} z 
        \left[ K_0 (P_E) I_1 (P_E z)
         +I_0 (P_E) K_1 (P_E z)\right] ,
\\
   d_1 (z)
     &\!\!\!=\!\!\!&
        {2 i \over \pi} z 
        \left[ K_0 (P_E z_L) I_1 (P_E z)
         +I_0 (P_E z_L) K_1 (P_E z)\right] ,
\\
   a_0 (z) 
     &\!\!\!=\!\!\!&
        {2\over \pi} z
          \left[
          K_0 (P_E) I_0 (P_E z)
            -I_0 (P_E) K_0 (P_E z) \right] ,
\\
  d_0 (z) &\!\!\!=\!\!\!&
        {2\over \pi} z
          \left[
          K_0 (P_E z_L) I_0 (P_E z)
            -I_0 (P_E z_L) K_0 (P_E z) \right] ,
\\
   a_2 (z) &\!\!\!=\!\!\!&
      {4 \over P_E \pi} \left[
       K_0 (P_E) I_1 (P_E z)
         +I_0 (P_E) K_1 (P_E z) \right]  -a_0 (z) ,
\\
  d_2 (z) &\!\!\!=\!\!\!&
      {4 \over P_E \pi} \left[
       K_0 (P_E z_L) I_1 (P_E z)
         +I_0 (P_E z_L) K_1 (P_E z) \right]  
   -d_0 (z) ,     
\eea
where Eqs.~(\ref{jy0}) and (\ref{jy1})
have been used.
With these equations,
$I_{G3}$ is written as Eq.~(\ref{IG3con}).

\section{Bessel function formula \label{app:bessel}}

The formula
\bea
  J_0 (P z) Y_1 (P z) - Y_0 (P z) J_1(P z) = -{2\over
   \pi P z} ,
\eea
is often used in the analysis in this paper.

Correspondingly to the Wick rotation,
the following equations are useful:
\bea
   J_0 (i z) &\!\!\!=\!\!\!& I_0 (z) ,
\qquad
   Y_0 (i z) =
      i I_0 (z) -{2\over \pi} K_0 (z) ,
      \label{jy0}
\\
   J_1 (i z) &\!\!\!=\!\!\!&
      i I_1 (z) ,
\qquad
   Y_1 (i z) =
      -I_1 (z) + {2 i \over \pi} K_1 (z) .
       \label{jy1}
\eea

\end{appendix}

\newpage



\end{document}